\documentclass[12pt,a4paper]{article}

\usepackage{graphicx} 
\usepackage{amsmath}
\usepackage{amssymb}
\usepackage{array} 
\usepackage[margin=1in]{geometry}
\usepackage{pdfpages}
\usepackage{svg}
\usepackage{float}
\providecommand{\keywords}[1]
{
  \small	
  \textbf{\textit{Keywords---}} #1
}

\title{SynerGraph: An Integrated Graph Convolution Network for Multimodal Recommendation}
\author{Mert Burak Burabak\footnote{Mert Burak Burabak\\
        Faculty of Engineering and Natural Sciences, Bahcesehir University\\
	Kadikoy, Istanbul, Turkey\\
	  E-mail: mbburabak@gmail.com},
  Tevfik Aytekin\footnote{Tevfik Aytekin\\
	Department of Computer Engineering, Bahcesehir University, Ciragan Caddesi, 34353\\
	Besiktas, Istanbul, Turkey\\
	Tel.: +90-212-3810580\\
	Fax: +90 212 3810550\\
	E-mail: tevfik.aytekin@bau.edu.tr\\
	\\}}

\begin{document}
\maketitle
\begin{abstract}
    This article presents a novel approach to multimodal recommendation systems, focusing on integrating and purifying multimodal data. Our methodology starts by developing a filter to remove noise from various types of data, making the recommendations more reliable. We studied the impact of top-K sparsification on different datasets, finding optimal values that strike a balance between underfitting and overfitting concerns. The study emphasizes the significant role of textual information compared to visual data in providing a deep understanding of items. We conducted sensitivity analyses to understand how different modalities and the use of purifier circle loss affect the efficiency of the model. The findings indicate that systems that incorporate multiple modalities perform better than those relying on just one modality. Our approach highlights the importance of modality purifiers in filtering out irrelevant data, ensuring that user preferences remain relevant. Models without modality purifiers showed reduced performance, emphasizing the need for effective integration of pre-extracted features. The proposed model, which includes an novel self supervised auxiliary task, shows promise in accurately capturing user preferences. The main goal of the fusion technique is to enhance the modeling of user preferences by combining knowledge with item information, utilizing sophisticated language models. Extensive experiments show that our model produces better results than the existing state-of-the-art multimodal recommendation systems.
    \\
    \\
\end{abstract}

\setcounter{page}{0}
\section{Introduction}

The rise of recommendation systems has significantly changed how we interact in the world customizing our experiences on various platforms by cleverly anticipating and suggesting options that match our individual tastes. Initially based on collaborative filtering and content based filtering these systems have evolved over time though they face challenges that have prompted the need for more advanced methods\cite{linden2003amazon, hu2008collaborative, rendle2009, su2009survey, lam2008addressing, koren2009}.
\newline
\newline
Traditional recommendation systems mainly use collaborative filtering, where algorithms predict what users might like based on their past interactions within a group of users. While effective these systems struggle with data and the cold start issue when dealing with new users or items with limited interaction history. Content based approaches, which recommend items to a users past preferences also have drawbacks in capturing the complex aspects of user interests and item characteristics. To tackle these challenges there has been a move towards multimodal recommendation systems. These systems combine types of data. Such, as text, images, sound and more. To develop a deeper understanding of user preferences and item features. This integration helps overcome the limitations of single mode data by providing a comprehensive view of user item interactions resulting in more precise and personalized recommendations. In this context Graph Neural Networks (GNNs) have emerged as a crucial technological development \cite{zhou2018graph, shi2015heterogeneous, hamilton2017inductive, kipf2017semi}. By representing data as a graph GNNs can grasp connections and interactions between users and items effectively utilizing the inherent structural information in the data. These networks excel at recognizing patterns and deducing user preferences from the web of user item interactions thereby significantly enhancing recommendation accuracy.
\newline
\newline
This study seeks to investigate the effectiveness of GNNs in recommendation systems by introducing a new framework that integrates different data types using a graph based approach. We explore how merging data modalities can overcome traditional system constraints and enhance the recommendation process. Our key contributions consist of; Introduction of a GNN based framework that efficiently handles multimodal data providing an in depth analysis of its structure, data processing techniques and integration mechanisms. Presentation of an evaluation of the proposed system, on various datasets emphasizing its predictive precision, adaptability and enhancements compared to existing approaches. Providing perspectives on how multimodal data fusion works in a graph based setting aiding in a better grasp of the processes that lead to improved recommendation results. Delving into the real world implications of our discoveries, such as uses, advantages for users and the broader influence on recommendation systems. Through this study the article aims to add a section to the ongoing story of recommendation system advancement laying groundwork for future exploration and progress, in utilizing GNNs for improved multimodal recommendations.
\section{Related Works}

In the changing world of recommendation systems incorporating multiple types of data has become a crucial strategy to overcome the limitations of traditional collaborative filtering methods. By blending textual and visual elements into recommendation frameworks significant progress has been made in understanding detailed user preferences and item attributes \cite{Liu2023MultimodalRS}.
\newline
\newline
A key player in this advancement is the Visual Bayesian Personalized Ranking (VBPR) \cite{He2015VBPRVB}. This approach combines features from item images with user item interactions to create a new way of understanding user tastes emphasizing the impact of visual appeal on user decision making. Expanding on this approach the Visual Semantic Embedding with Co Attention Fusion (VECF) \cite{Feng2021EncoderFN} represents a substantial step forward. VECF utilizes both textual information using a co attention mechanism to form a unified representation that clarifies the intricate connection between user preferences and item qualities. The introduction of Graph Neural Networks (GNNs) has further transformed recommendations. MMGCN \cite{Hu2021MMGCNMF} utilizes a modality structure based on user item interactions, within a graph framework showcasing how GNNs can effectively process complex multimodal data. In a vein the GRCN model \cite{Wei2020GraphRefinedCN} introduces a new method for refining graphs that carefully organizes the connections between users and items to reduce errors and unrelated correlations typically found in large datasets thereby improving recommendations. Taking an approach the DualGNN \cite{Wang2023DualGNNDG} utilizes graph structures to capture relationships between users and items providing a nuanced view of how they interact that traditional models often miss. The LATTICE framework \cite{Zhang2021MiningLS} emphasizes exploring structures in multimodal recommendation systems. By creating and merging item item graphs from perspectives LATTICE delves into the underlying connections among items enhancing our understanding of their relationships. In addition SLMRec \cite{Wei2023MultiModalSL} combines self supervised learning with GNNs to uncover hidden patterns in data. This fresh strategy not only improves how data is represented but also opens up possibilities for using unsupervised cues to enhance recommendation systems. Lastly BM3 \cite{Zhou2022BootstrapLR} highlights the importance of refining latent representations by blending ID embeddings with features to address issues, like sparse data and noise effectively a strategic fusion of diverse insights and representation learning.
\newline
\newline
These advancements together represent a change, in recommendation systems highlighting the powerful influence of combining different types of data and using graph based learning to create more intuitive, flexible and user focused recommendation engines.
\section{Methodology}

\subsection{Multimodal Graph Construction}

\subsubsection{Overview of the Amazon Reviews Dataset}
The dataset known as Amazon Reviews is a resource, for studying data in recommendation systems. It consists of a variety of customer feedback including both written reviews and product images. In our research we specifically focus on three categories; toys, sports equipment and office supplies. These categories were chosen because they offer characteristics and diverse types of reviews and images\cite{amazon2014dataset}.

\subsubsection{Textual Data}
The textual part of the dataset includes user generated reviews and product descriptions which provide insights into user preferences, opinions and the features of products. The reviews range from comments to narratives representing the experiences and viewpoints of users. On the hand product descriptions information from the manufacturers perspective highlighting important features and specifications.

\subsubsection{Image Data}
The inclusion of images in this dataset complements the information by offering a representation of products. Images capture aspects such as aesthetics and functionality that may not be fully conveyed through text alone. For example visual cues like toy colors, designs and sizes; sports equipment quality and style; or ergonomic design in office supplies can greatly influence user preferences when making purchasing decisions.

\subsubsection{Significance of Multimodality}
By combining both text based information and visual data from images recommendation systems can gain a understanding of products as well, as user preferences.
This approach of combining modes of data is beneficial as it captures the details of how users interact with products, which are often overlooked in systems that consider only one mode.

\subsection{Node and Edge Definition}

In this section we will explain the structure and design of nodes and edges, in our recommendation system, which is based on a graph. Our focus is on how we represent products as nodes by utilizing both generated image data and textual data from the dataset. We will also discuss how connections between products are established without attributions.

\subsubsection{Product Node Representation}
\begin{itemize}
    \item \textbf{Utilization of Pre-Generated Image Features:} Each product node incorporates predefined image features provided by the dataset owners. These features capture aspects such as color, design and layout\cite{amazon2014dataset}.
    \item \textbf{Textual Feature Integration:} In addition to features we process data from product descriptions using sentence transformers to create semantic feature vectors. This process captures qualitative information about the products \cite{Reimers2019SentenceBERTSE}.
\end{itemize}

\subsubsection*{Edge Definition Based on Historical Purchases}
\begin{itemize}
    \item \textbf{Utilizing Historical Purchase Data:} Edges within our graph are determined based on purchase data from the Amazon Reviews dataset.
    \item \textbf{Graph Topology and Consumer Insight:} This approach allows our graph to depict a network of products based on real consumer behaviors. It helps in generating recommendations that reflect the buying habits of people.
\end{itemize}

\subsubsection{Implications of the Product-Centric Graph Structure}
The graph design, which focuses on product centered nodes and simple connections showcases our strategy of utilizing the characteristics of the products. By combining generated image features, with data we structure the graph in a way that allows us to extract and analyze patterns based on a comprehensive representation of the products. This lays the groundwork for a recommendation system.

\subsection{Feature Extraction}

In this section we explain the methods used to extract and process features from two types of data in our dataset; text and images. The extraction process is crucial for converting data into a format for analysis using graphs.

\subsubsection{Textual Feature Extraction}
\begin{itemize}
    \item \textbf{Employment of Sentence Transformers:} To extract features from text based data like product descriptions and titles we employ sentence transformers. These transformers utilize techniques of natural language processing to convert text into dimensional semantic feature vectors\cite{Reimers2019SentenceBERTSE}.
    \item \textbf{Semantic Understanding:} The sentence transformers are designed to capture meanings and nuances of words ensuring that the derived features accurately represent the content of the text.
    \item \textbf{Vectorization Process:} Each piece of data goes through tokenization embedding and aggregation processes, within these transformers to create a feature representation.
\end{itemize}

\subsubsection{Image Feature Extraction}
\begin{itemize}
    \item \textbf{Utilization of Pre-Generated Features:} The dataset contains image features that were generated in advance for each product. These features were obtained by applying image processing algorithms as mentioned by the dataset providers\cite{amazon2014dataset}.
    \item \textbf{Feature Characteristics:} These image features effectively capture aspects such, as color, texture, shape and pattern. They provide a representation of the appeal of each product.
    \item \textbf{Integration with Textual Features:} The pre generated image features are used as they are and combined with feature vectors to create a feature set for every product node in the graph. This integration allows the system to utilize both textual information for recommendation purposes.
\end{itemize}

\subsubsection{Importance of Feature Extraction}
\begin{itemize}
    \item \textbf{Foundation for Analysis:} The extracted features serve as the foundation for graph based analysis. It is crucial to extract features to ensure the effectiveness of the recommendation system.
    \item \textbf{Enabling Multimodal Insights:} By combining text and image features the system can develop a understanding of each product resulting in more precise and relevant recommendations.
\end{itemize}

\subsection*{Graph Representation}

The way we represent the graph plays a role in the effectiveness and efficiency of our recommendation system. In this section we explore why we chose a format for graph representation and discuss its implications.

\subsubsection*{Utilizing Adjacency Matrix for Graph Structure}
\begin{itemize}
    \item \textbf{Matrix Structure Adoption:} We structure the graph using an adjacency matrix, where rows and columns represent nodes and each matrix element indicates whether there is an edge between pairs of nodes. We chose this format because it provides a representation of the connections within the graph.
    \item \textbf{Advantages for Graph Algorithms:} The adjacency matrix format is well suited for graph neural network algorithms as it allows for implementation of operations, like neighborhood aggregation or spectral convolutions.
\end{itemize}

\subsubsection*{Optimization Strategies for Sparse Graphs}
\begin{itemize}
    \item \textbf{Sparse Matrix Representation:}Considering the graphs expected sparsity, which means there are edges compared to the number of possible connections we use a sparse matrix representation. This approach allows us to focus on storing and processing the zero elements, which brings significant memory and computational benefits.
    \item \textbf{Enhancing Computational Performance:} Using matrices is especially advantageous when dealing with large scale datasets. It greatly reduces the workload during operations, like matrix multiplication, which's a common step in graph based learning algorithms.
\end{itemize}

\subsubsection*{Facilitating Advanced Graph-Based Learning}
\begin{itemize}
    \item \textbf{Supporting Complex Algorithms:} Choosing an adjacency matrix in its form plays a crucial role in enabling the application of advanced graph based learning algorithms. This includes types of graph networks that are essential, for learning from and making predictions based on integrated multimodal data.
\end{itemize}

\section{Model Architecture Design}
In this part we'll introduce the structure of the proposed model. The foundation of the model is built on a framework that defines users and items in an embedding space. This space helps enhance our understanding of interactions and inferences. The design of the model approaches the problem from angles providing a range of functions for different tasks. Firstly we will give the problem definition and discuss how we refine modalities. Then we'll explain how various types of interactions are processed within the model, including user item and item item interactions across modalities. Following this we'll introduce how we combine the information gathered from each modality and present the information fusion layer. After these we will outline our additional task, loss functions and training process.

\begin{figure}[t]
    \centering
    \caption{Proposed Model Architecture}
    \label{Proposed Model Architecture}
    \includegraphics[width=1\linewidth]{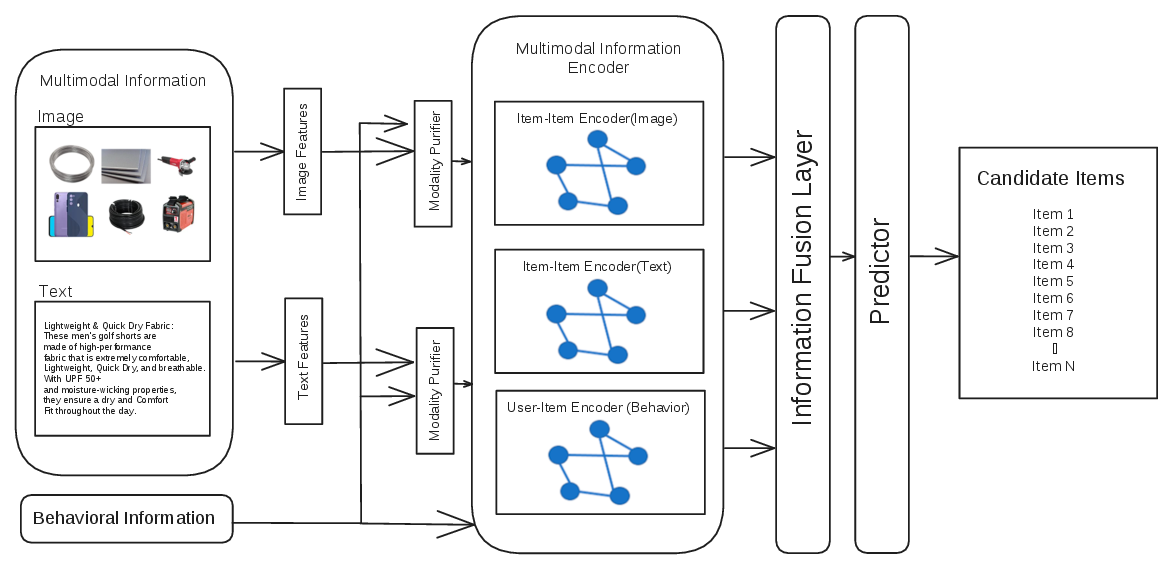}
\end{figure}

\subsection{Problem Definition}
Now lets define a set called $U$ that consists of users represented by elements $u$ and another set called $I$ comprising items denoted by elements $i$.In our approach we assign a vector representation, to each user and item in a Euclidean space called the embedding space. This space has a dimensionality of $d$.

For each item we recognize the existence of feature representations indexed as $E_{m i}$ in a space with dimensionality $d_m$. These feature representations come from the modality set $M$, where for this discussion we focus on the visual and textual characteristics represented by $M = \{v, t\}$.

It's important to note that although our discussion is limited to these modalities the framework we propose is not restricted and can be extended to include a wider range of modalities.

Next we introduce a dataset called interactions denoted as $R$. This dataset represents relationships within the space $\{0, 1\}^{|U| \times |I|}$. Each element $R_{u i}$ is set to 1 if there was an interaction, between user $u$ and item $i$ and 0 otherwise.We can think of this historical interaction dataset, as a graph, where users and items are represented as nodes.

This historical interaction dataset can be conceptualized as a sparse bipartite graph $G = (\gamma, \mathcal{E})$, with $\gamma$ being the union set $\{U \cup I\}$ representing nodes corresponding to users and items, and $\mathcal{E}$ being the set of edges, constituted by pairs $(u,i)$ such that $R_{u,i} = 1$, illustrating the interaction history.

Our goal is to create a solution for a multimedia recommendation system. This system should accurately. Rank items, for users based on their predicted preference scores, which we denote as $y_{u i}$.

\subsection{Multi Modality Handler}
The effectiveness of multimedia recommendations is greatly influenced by both signals and related cues. As a result we have embraced the approach presented in \cite{Zhang2021MiningLS}. Developed a view information encoder to enhance the distinctiveness of features. This encoder collects signals by analyzing interactions, between users and items well as semantically related signals by examining the relationships between different items.

\subsubsection{Modality Purifier}

The process begins by transforming raw modality features referred to as $E_m$ into higher level representations called $\tilde{E_m}$. This transformation is achieved using a matrix $W$ and a bias vector $b$;

\begin{equation}
    \tilde{E}_m = W E_m + b,
\end{equation}

In this case $W$ is a matrix with dimensions $d \times m$, where $d$ represents the desired dimensionality of the higher level features. The bias vector $b$ has dimensions of $d$. These parameters can be adjusted during training to optimize the feature representations.

To extract modality features to user preferences from the higher level representations ${G_m}$ we introduce a gating mechanism.

\begin{equation}
    {G_m} = {G_i} \odot \sigma(W \tilde{E_m} + b),
\end{equation}

In this equation:
\begin{itemize}
    \item $\odot$ represents element-wise multiplication.
    \item $\sigma$ denotes the tanh function.
    \item $W$ is a trainable matrix of dimensions $d \times d$, and $b$ is a learnable bias vector.
    \item ${G_i}$ represents initial learnable embedding of user preferences.
    \item $\tilde{E_m}^{cm}$ represents the refined modality features, specifically chosen to align with user preferences.
\end{itemize}

 This mechanism uses user features encoded in the embedding ${G_i}$ to control the impact of modality features in $\tilde{E_m}$. By doing it focuses on enhancing and isolating the modality components that're most important for user preferences. The resulting refined representation of item modality features ${G_m}$, captures details of user item interactions. Ultimately improves recommendation quality.

\subsubsection{User - item Interactions}
The main goal of this encoder is to distill signals by encapsulating higher order relational dynamics within the user item interaction graph. At each level of the graph network (GCN) we apply a message propagation protocol to enhance representations of users and items \cite{he2020lightgcn}. To provide an analysis of how user item interactions evolve we propose the following process;

\begin{equation}
    G_{i}^{(l)} = G_{i}^{(l-1)} \cdot L
\end{equation}

In this equation \(G_{i}^{(l)}\) represents advanced representations at the \(l\) th layer, \(G_{i}^{(0)}\) denotes initial identity embeddings and \(L\) corresponds to the Laplacian matrix derived from the user item graph;

\begin{equation}
    L = D^{-\frac{1}{2}}\tilde{A}D^{-\frac{1}{2}},
\end{equation}, \begin{equation}
    \hat{A} = 
    \begin{bmatrix}
    0 & R \\
    R^T & 0
    \end{bmatrix},
\end{equation}

Here \(\tilde{A}\) signifies an adjacency matrix, \(R\) represents the user item interaction matrix and \(D\) stands for the diagonal degree matrix. The normalization step ensures that embeddings maintain a scale throughout iterations.

By capturing neighborhood information up to connections our encoder contextualizes user item relationships effectively. The aggregated representation \(\tilde{G}_{i}\) is computed as an average, across all layers;

\begin{equation}
    \tilde{G}_{i} = \frac{1}{L + 1} \sum_{i=0}^{L} G_{i}^{(i)}
\end{equation}

This technique improves filtering by incorporating interaction signals.

\subsubsection{Item - Item Relationships}
Graph convolutions are used to item item relationships. Capture semantically correlated signals, with precision.
This process not improves the recommendation systems ability to predict by using both textual features but it also ensures that complex modality graphs are seamlessly integrated into an analytical framework. We start by measuring the connections, between items within each modality creating an interconnected graph. The resulting matrix, denoted as \( S_m \) contains weights that represent how similar the feature vectors of items are:
\begin{equation}
    s_{ij}^m = \frac{\mathbf{g}_i^{m^\top} \mathbf{g}_j^m}{\| \mathbf{g}_i^m \|_2 \| \mathbf{g}_j^m \|_2}
\end{equation}
where \( \mathbf{g}_i^m \) and \( \mathbf{g}_j^m \) represent the feature vectors of items \( i \) and \( j \) in modality \( m \), and \( s_{ij}^m \) is the corresponding normalized similarity metric.

To enhance the importance of connections we utilize graph convolution operations that filter and retain the most significant ones. This is achieved through a top-K(KNN) sparsification mechanism \cite{Shi2019UnderstandingTS} in which we focus on the most relevant features based on affinity scores.
\begin{equation}
    w_{ij}^m = 
    \begin{cases} 
    s_{ij}^m & \text{if } s_{ij}^m \text{ is in the top } K \text{ of } S_m \\
    0 & \text{otherwise}
    \end{cases}
\end{equation}
To maintain stability during learning and prevent any issues, with exploding gradients we normalize using a degree matrix \( D^m \). This matrix is calculated as the matrix of edge weight sums.
\begin{equation}
    D^m = \text{diag}(\sum_j s_{ij}^m)
\end{equation}
The item affinity matrix \( E_m \) is updated through a Graph Convolutional Network (GCN) to propagate features effectively.
\begin{equation}
    G_m = D^{m^{-\frac{1}{2}}} S_m D^{m^{-\frac{1}{2}}} \cdot G_m
\end{equation}

To create a representation we focus on the common characteristics shared by different items. We utilize the structure of the GCN (Graph Convolutional Network) to integrate these features gradually increasing their depth:
\begin{equation}
    G_m^{(l+1)} = \sigma\left(\hat{D}^{m^{-\frac{1}{2}}} \hat{S}_m \hat{D}^{m^{-\frac{1}{2}}} G_m^{(l)}\right)
\end{equation}
The activation function \( \sigma \) introduces non linearity into the process of integrating these features. Eventually we normalize the combined features in order to maintain their distinctiveness and prevent dilution, across modalities using the multimodal feature matrix \( G_m \):
\begin{equation}
    G_m = \frac{G_m}{\sqrt{\| G_m \|_F}}
\end{equation}
where \( \| \cdot \|_F \) represents the Frobenius norm, ensuring a uniform scale for the aggregated feature set.

\subsubsection{Fusion Layer}
To transform user behavior features into modality preferences we employ a transformation that considers parameters and utilizes a tanh activation function. This helps model the interactions between user behavior and modality features:
\begin{equation}
    G_m = \sigma(W G_{i} + b),
\end{equation}

In this transformation we use \( \sigma \) as the tanh function while \( W\in, \mathbb{R}^{d \times d} \). \( b\in \mathbb{R}^d\) represent parameters. The variable \( G_i\) represents user behavior features.

We ensure a mapping of user attention by employing an attention mechanism to extract shared features across modalities:
\begin{equation}
    a_m = \text{softmax}\left(\mathbf{q}^\top \tanh(W G_m + b)\right),
\end{equation}

The learned attention vector \( \mathbf{q}\) weight matrix \( W\) and bias vector \( b\) are all shared across all modalities to reflect the components of user interaction.

The features that are shared across modalities are combined to create a set of features that represent the preferences of users, across all modalities:
\begin{equation}
    G_s = \sum_{m \in M} a_m G_m,
\end{equation}
In this formula \( G_s \) represents the combined feature set and \( M \) represents all the modalities. This combined set is believed to improve the accuracy of the recommendation system.

To generate the user and item representations we use the equation;

\begin{equation}
    G_i = G_i + \frac{G_s}{\sqrt{\| G_s \|}}
\end{equation}

\subsection{Loss Function}
This section explains how a comprehensive loss function is developed for a recommendation system that deals with multiple modes of data. The function combines loss components with an use of circle loss to overcome challenges associated with multimodal data.

\subsubsection{BPR Loss Formulation}
The BPR loss \cite{rendle2009}, referred to as \(L_{BPR}\) is specifically designed to optimize the order in which users interact with items based on their representations.The formula is expressed as;
\begin{equation}
    L_{BPR} = -\sum_{(u, i, j) \in \Omega} \log \sigma(\mathbf{u}_g \cdot \mathbf{i}_{g,pos} - \mathbf{u}_g \cdot \mathbf{i}_{g,neg}),
\end{equation}
where \( \Omega \) represents the collection of user item combinations \( \mathbf{u}_g \) stands for the user representation, \( \mathbf{i}_{g,pos} \) and \( \mathbf{i}_{g,neg} \) are the representations of negative items and \( \sigma \) symbolizes the sigmoid function.

\subsubsection{Embedding Loss}
The embedding loss, denoted as \(L_{Emb}\), aims at regularizing the learned embeddings:
\begin{equation}
    L_{Emb} = \lambda (\| \mathbf{U} \|^2_F + \| \mathbf{I_p} \|^2_F + \| \mathbf{I_n} \|^2_F),
\end{equation}
where \( \mathbf{U} \), \( \mathbf{I_p} \), \( \mathbf{I_n} \) are user, positive and negative item embedding matrices, respectively, and \( \lambda \) is the regularization coefficient.

\subsubsection{Circle Loss Integration for Multimodal Fusion}
In recommendation systems accurately representing user preferences can be difficult due to modality noise. To overcome this challenge, a modified version of the Circle Loss function \cite{Sun2020CircleLA} is introduced. This enhanced function handles sample variability and modality noise more effectively. The critical interpretation and modification of the Circle Loss in this context aim to achieve the following objective: even though the presence of individual modalities may be reduced through purifier processes, they still exist. Therefore, the goal is to minimize data contamination found in individual modalities when creating a fused modality. So, we tried to discriminate fused modality from individual modalities.

Positive similarity is calculated as:
\begin{equation}
S_{pos} = \cos(u, i_{fused})
\end{equation}
Negative similarity is calculated as:
\begin{equation}
S_{neg} = \cos(u, i_{modality})
\end{equation}

where $u$ is user embedding, $i_{fused}$ is positive item's fused embedding and $i_{modality}$ is positive item indivudual modality embedding.

The similarities are adjusted using a margin parameter \(M\):
\begin{align}
AP &= \max(-S_{pos} + 1 + M, 0) \\
AN &= \max(S_{neg} + M, 0)
\end{align}

If negative confidence \(C_{neg}\) is provided, the negative similarity is scaled:
\begin{equation}
AN = AN \times (1 - C_{neg})
\end{equation}
where \(C_{neg}\) is confidence level representation of embedding accuracy.

The logits are computed as:
\begin{align}
L_{pos} &= -AP \times (S_{pos} - \Delta_{p}) \times \Gamma \\
L_{neg} &= AN \times (S_{neg} - \Delta_{n}) \times \Gamma
\end{align}

where $\Delta_{p} = 1 - M$, $\Delta_{n} = M$ and $\Gamma$ amplify the differences between the adjusted similarities for positive and negative samples

The final Circle Loss is calculated over all samples as:
\begin{equation}
L = \text{mean}\left( \text{softplus}\left(\log\sum\exp(L_{pos}) + \log\sum\exp(L_{neg})\right)\right)
\end{equation}

\textbf{Advantages in Recommendation Systems}
\begin{itemize}
    \item \textbf{Effective Modality Noise Mitigation}: By incorporating a mechanism that adjusts the impact of reliable similarities the model can effectively consider different modalities and adapt its weighting accordingly.

    \item \textbf{Enhanced Discriminative Learning}: The combination of margin based transformation and logit computation ensures a distinction between noisy individual modalities and fused final modality within the feature space promoting discriminative features.

    \item \textbf{Dynamic Adaptability}: The design of the Circle Loss function allows it to flexibly adapt to levels of reliability in item embeddings making it highly effective, in various recommendation scenarios.
\end{itemize}

\subsubsection{Total Loss Calculation and Optimization Procedure}
The total loss, \(L_{Total}\), is an aggregation of the matrix factorization loss, embedding loss, and modified circle loss:
\begin{equation}
    L_{Total} = L_{BPR} + L_{Emb} + \gamma \cdot (L_{Circle(text)} + L_{Circle(image)}),
\end{equation}
where \( \gamma \) is the weight factor for the circle loss.

During the optimization process various techniques such, as zeroing, backpropagation, gradient clipping and parameter updating are used.

\subsection{Training Procedure}

One crucial aspect of our algorithm is the training procedure. It plays a role in optimizing the model parameters to improve recommendation accuracy. Our training process consists of epochs where each epoch represents an iteration through the training dataset. Within each epoch we carefully execute the following operations;

\begin{enumerate}
    \item \textbf{Random Negative Sampling:}  We incorporate sampling to enhance the training process. Along with user item interactions we randomly sample item IDs to ensure that our model encounters a diverse range of user interactions.
    
    \item \textbf{Batch Sampling:} We use a sampling strategy to select both negative user item pairs. This approach allows our model to be exposed to a combination of user interactions.
    
    \item \textbf{Embedding Computation:} We compute embeddings for users, items and multimodal fusion which serve as the foundation for computations.
    
    \item \textbf{Loss Computation:} Our training loop incorporates components for loss calculation including matrix factorization, embeddings and Circle Loss. Each with its weighting.
    
    \item \textbf{Backpropagation and Optimization:} During backpropagation gradients of the batch loss with respect, to the models parameters are computed.These gradients are instrumental, in optimizing parameters, which ultimately results in recommendations. We utilize AdamW \cite{Loshchilov2017FixingWD} as the optimizer.
\end{enumerate}
\section{Experimentation and Results}

In this chapter we will explore the process of conducting experiments. The outcomes achieved through our multimodal recommendation system. Lets begin by discussing the datasets we utilized with a focus, on the 5 core configuration and the careful division of items into training, validation and test sets. Next we will outline the steps taken to preprocess the data.

\subsection{Dataset Selection}

For our experiments we made selections from three datasets on Amazons platform. These datasets were specifically chosen in the 5 core setup to ensure that each user and item in them had 5 interactions \cite{amazon2014dataset}. Our aim in this selection process was to obtain comprehensive interaction data for our recommendation system. Here are the datasets we chose;

\begin{enumerate}
    \item \textbf{Amazon Baby}: This dataset includes a range of baby products, like diapers, strollers, clothing and feeding accessories.
    \item \textbf{Amazon Sports}: The dataset contains interactions between customers and sports and outdoor products on the platform. It gives us a view of consumer preferences and trends in sports through items such as apparel and sporting equipment.
    \item \textbf{Amazon Clothing}: This dataset presents an overview of how customers engage with clothing items on Amazon. It covers types of apparel from wear to formal outfits reflecting the diverse tastes and buying habits of fashion conscious Amazon shoppers.

\end{enumerate}

These datasets cover a range of product domains, which adds richness to the variety of user item interactions in our experiments.

\subsection{Data Preprocessing}

The preprocessing pipeline for our data focused on preparing the datasets for model training and evaluation. Here is a detailed overview;

\begin{itemize}
    \item \textbf{User and Item Encoding:} To facilitate handling of user item interactions within the model during training and evaluation we encoded users and items with identifiers.
    
    \item \textbf{Text Features:} Textual data was processed using Sentence Transformers (all-mpnet-base-v2) \cite{Reimers2019SentenceBERTSE}.
    
    \item \textbf{Image Features:} When it came to image data like product images we directly extracted them from the dataset itself \cite{McAuley2015ImageBasedRO}. This eliminated the need, for image processing while providing visual information to our model.
\end{itemize}

\subsection{Dataset Statistics}
The table presents a comparative overview of three distinct datasets: Baby, Sports, and Clothing. Each dataset is evaluated based on four criteria: number of users, number of items, number of interactions, and sparsity.

\begin{table}[H]
\centering
\begin{tabular}{lcccc}
\hline
\textbf{Dataset} & \textbf{\#(Users)} & \textbf{\#(Items)} & \textbf{\#(Interactions)} & \textbf{Sparsity} \\ \hline
Baby             & 19,445             & 7,050              & 160,792                  & 99.883\%         \\
Sports           & 35,598             & 18,357             & 296,337                  & 99.955\%         \\
Clothing         & 39,387             & 23,033             & 278,677                  & 99.969\%         \\ \hline
\end{tabular}
\caption{Overview of the Baby, Sports, and Clothing datasets.}
\label{tab:dataset_overview}
\end{table}

By examining these statistics we can gain an understanding of the scale of the datasets and how users interact with items. These insights play a role in framing our results and analyzing the performance of our multimodal recommendation system.

\subsection{Train-Test Split}

One crucial aspect of preparing our dataset was ensuring a train test split. To ensure model training and evaluation we followed this approach;

\begin{itemize}
    \item \textbf{Random Split:} Items, within each dataset were randomly partitioned into three sets; training (80\%) validation (10\%) and test (10\%). This randomization helped prevent biases in how itemsre distributed across these sets.

    \item \textbf{User-Centric Split:} To ensure that we effectively utilized each users interactions, for both training and evaluation purposes we made sure that all interactions from an user were included in either the training, validation or test set. This approach allowed us to create datasets that were centered around the users.
    
    \item \textbf{Balanced Distribution:} We strived to maintain a distribution of data across users ensuring that users with varying levels of activity were represented in all sets. This helped prevent any biases in recommendations caused by an overrepresentation of users.
\end{itemize}

Our careful methodology for splitting the data into training and testing sets ensured that our model was trained on data while also being rigorously evaluated on interactions it had never seen before.

These careful preprocessing steps were undertaken to optimize the datasets, for training and evaluating our multimodal recommendation system.

\subsection{Evaluation Metrics}

\textbf{Definition:} When it comes to evaluating recommendation models, fairness in evaluation metrics necessitates using a set of criteria. All models should be assessed using standards.

\textbf{Implementation:} Select a standard set of evaluation metrics and apply them uniformly to evaluate all models. Here are the key metrics with their formulas:

\begin{itemize}
    \item \textbf{Recall at 20 (Recall@20):}
    
    \[ \text{Recall@20} = \frac{\text{Number of Relevant Items in the Top 20 Recommendations}}{\text{Total Number of Relevant Items}} \]

    \item \textbf{Normalized Discounted Cumulative Gain at 20 (NDCG@20):}
    
    \[ \text{NDCG@20} = \frac{1}{\log_2(1 + \text{Rank}_1)} \times \frac{1}{\log_2(1 + \text{Rank}_2)} \times \ldots \times \frac{1}{\log_2(1 + \text{Rank}_{20})} \]
\end{itemize}

By utilizing these metrics we ensure that each models performance is assessed against the benchmarks enabling comparisons.

\subsection{Compared Methods}
To evaluate the performance of recommendation models we compared cutting edge methods that each represent an approach to recommendation. For implementations of compared methods, the implementations of paper is used \cite{malitesta2023formalizing}.The methods we compared are;

\begin{enumerate}
    \item \textbf{ItemKNN:}
     This method calculates item similarities and assesses how similar a set of items is to a recommended item. It utilizes these relationships between items to generate recommendations\cite{Deshpande2004ItembasedTR}.

    \item \textbf{Matrix Factorization (MF):}
     This optimization technique leverages the Bayesian Personalized Ranking (BPR) loss. It focuses on modeling interactions between users and items as target values, for the interaction function\cite{rendle2009}.

    \item \textbf{Neural Graph Collaborative Filtering (NGCF):}
    this method explicitly models user item interactions using a graph structure.By incorporating graph operations NGCF allows for the interaction, between users and items embeddings. This enables the capturing of both signals and high order connectivity signals\cite{wang2019neural}.

    \item \textbf{LightGCN:}
    LightGCN challenges the complexity of Graph Convolutional Networks (GCNs) used in recommendation systems\cite{he2020lightgcn}.

    \item \textbf{Visual Bayesian Personalized Ranking (VBPR):}
    VBPR extends the BPR model by incorporating features and item ID embeddings to represent each item. In our experiments we combine modal features to predict user item interactions within a Matrix Factorization framework\cite{He2015VBPRVB}.

    \item \textbf{Multimodal Graph Convolutional Network (MMGCN):}
    MMGCN work on modal recommendation systems constructs modal specific graphs and refines modal specific representations for users and items. It then aggregates these representations to obtain comprehensive user and item representations for prediction\cite{Hu2021MMGCNMF}.

    \item \textbf{Graph Recurrent Collaborative Network (GRCN):}
    GRCN refines the user item interaction graph by identifying false positive feedback and removing corresponding noisy edges.bThis process improves the quality of the interaction graph to ensure recommendations\cite{Wei2020GraphRefinedCN}.

    \item \textbf{LATTICE: } 
    LATTICE combines modality graphs with the user item graph to extract comprehensive semantics for collaborative filtering\cite{Zhang2021MiningLS}.
    
\end{enumerate}

By evaluating these methods on our chosen datasets our aim is to gain insights into how they perform compared to each other and identify the strengths and weaknesses of recommendation approaches.

\subsection{Hyper parameter Tuning}
Optimizing the performance of recommendation models involves a step called hyperparameter tuning. It includes exploring hyperparameters to find configurations that produce the best results. The choice of hyperparameters can greatly impact how well the model captures user preferences and generates recommendations.

In our experiments we carefully selected a range of hyperparameters for each recommendation model. We considered factors such as learning rates, regularization strengths embedding dimensions and batch sizes. To efficiently explore the hyperparameter space we utilized a technique called parameter space search introduced by \cite{Akiba2019OptunaAN} in their work on Optuna.

The goal of hyperparameter tuning is to strike a balance between model complexity and generalization. Overfitting occurs when a model performs well on training data but poorly, on data. By making choices in hyperparameters we can mitigate overfitting. Achieve better generalization.

On the hand when the model lacks the ability to capture patterns it is referred to, as underfitting. To address this issue we can increase the complexity of the model by adjusting hyperparameters.

Our approach to tuning hyperparameters was based on evaluation and validation using a validation set. We adjusted hyperparameters, trained models and evaluated their performance using appropriate metrics.

The specific hyperparameter values we selected played a role in achieving results as presented in the following sections. The process of tuning hyperparameters ensures that our models are fine tuned to suit the characteristics of the datasets and can provide recommendations to users.

For settings we kept the embedding size fixed at 64 batch size at 1024 epochs at 200 learning rate decay at 0.00001 and random seeds at 123. We used Adam optimizer for optimizing all models. In terms of filtering (CF) models, in general we explored a learning rate range of (0.0001, 0.0005, 0.001 0.005). For models specifically we utilized the hyperparameter settings reported in the original baseline papers.

\subsection{Experiment Results}

In this section we will compare the approach we developed with the baseline. We will use the Recall@20. NDCG@20 metrics to conduct these comparisons. For our proposed approach, following hyper parameters of circle loss are used: 
\begin{itemize}
    \item circle loss co-efficient: 0.1
    \item margin = 0.75
    \item gamma = 1000
    \item text embeddings confidence level = 0.7
    \item image embeddings confidence level = 0.3
\end{itemize}

You can find the results of our experiments, in Table 2. Here are the findings we obtained from analyzing the experiment outcomes;

\begin{enumerate}
    \item \textbf{Proposed approach significantly outperforms baseline models.} Our approach is specifically designed for providing recommendations in multimedia content. With directly utilizing information about the different types of media we enhance it by incorporating user behavior data to filter out any irrelevant disturbances. The process involves an encoder that captures collaborative and semantic relationships between items refining both behavior and modality features. By combining these features based on each users preferences we create profiles for users and items using a modality fuser. Moreover we employ a self supervised task that strengthens the connection between user characteristics and the combined multimodal features while reducing any noise from modalities. This approach allows our model to accurately identify user preferences across features without relying solely on simple cues, from behavior or multimodal data. As a result our method outperforms existing techniques across three datasets.

\begin{table}[ht]
\centering
\begin{tabular}{lcccccc}
\hline
\textbf{Model} & \multicolumn{2}{c}{\textbf{Clothing}} & \multicolumn{2}{c}{\textbf{Baby}} & \multicolumn{2}{c}{\textbf{Sports}} \\
 & Recall@20 & NDCG@20 & Recall@20 & NDCG@20 & Recall@20 & NDCG@20 \\
\hline
ItemKNN & 0.0275 & 0.0131 & 0.0317 & 0.0152 & 0.0403 & 0.0206 \\
BPRMF & 0.0200 & 0.0086 & 0.0430 & 0.0179 & 0.0440 & 0.0202 \\
LightGCN & 0.0443 & 0.0196 & 0.0729 & 0.0304 & 0.0807 & 0.0358 \\
NGCF & 0.0327 & 0.0139 & 0.0543 & 0.0218 & 0.0686 & 0.0293 \\
VBPR & 0.0501 & 0.0226 & 0.0680 & 0.0293 & 0.0847 & 0.0371 \\
MMGCN & 0.0306 & 0.0129 & 0.0474 & 0.0199 & 0.0535 & 0.0223 \\
GRCN & 0.0476 & 0.0201 & 0.0599 & 0.0259 & 0.0616 & 0.0267 \\
LATTICE & 0.0681 & 0.0301 & 0.0788 & 0.0347 & 0.0910 & 0.0399 \\
\textbf{SynerGraph} & \textbf{0.0840} & \textbf{0.0381} & \textbf{0.0866} & \textbf{0.0367} & \textbf{0.1036} & \textbf{0.0453} \\
\hline
\textbf{Improv.} & 23.30\% & 26.56\% & 9.92\% & 5.90\% & 13.86\% & 13.35\% \\
\textbf{p-value} & \textless0.05 & \textless0.05 & \textless0.05 & \textless0.05 & \textless0.05 & \textless0.05 \\
\hline
\end{tabular}
\caption{Comparison of recommendation models across different categories.}
\label{tab:model_comparison}
\end{table}

    \item \textbf{GCN based techniques can be affected by interference caused by modality noise.} Specifically VBPR outperforms MF by integrating behavioral features while MMGCN falls short compared to LightGCN. The main reason for this is the message propagation mechanism, in GCN based methods, which results in the widespread distribution of noise and subsequent degradation of user and item representations. On the hand our approach introduces a behavior guided purifier that effectively filters out modality noise right from the beginning. Additionally we have implemented a revised circle loss technique to further reduce modality noise. In our framework we assign embedding confidence levels to modality embeddings. Used these levels to eliminate noise originating from individual modalities when they are combined. As a result our approach has achieved performance in terms of recommendations.

    \item \textbf{To effectively address the problem of modality noise contamination it may be more beneficial to take an indirect approach by integrating modality features.} With directly including modality features in representations the GRCN technique utilizes these features exclusively to refine the user item interaction graph. Similarly LATTICE unveils concealed item relationships in order to create additional item item graphs for message propagation. However both GRCN and LATTICE encounter a challenge. They fail to incorporate modality information into the learning process of representations thus limiting their ability to fully capture user preferences. In contrast our method tackles the issue of contamination with a strategy; utilizing a purifier and circle loss. By purifying the modality features we enhance them for a nuanced integration, with behavior features. This refined fusion process ultimately leads to improved outcomes.

\end{enumerate}

\section{Discussion}
This chapter continues from the section, "Experimentation and Results." Here we have observed that the proposed approach has been successful based on the findings from the preceding section. We will now conduct analyses on the proposed approach.

\subsection{Effects of top-K Sparsification}
The embeddings obtained through transfer learning for both text and image modalities are not perfect. They contain some elements. As we saw in the results section directly using these embeddings in the model leads to attempts. Additionally one of the layers that make these embeddings useful is the top k sparsification layer. The parameter K in this layer helps determine which neighbors are closest to each embedding. The model learns from information about these K neighbors and calculates model weights accordingly. Setting a value for K can result in underfitting while a high value may cause overfitting. Hence we have conducted complementary analyses on our datasets to determine the optimal K value, for each dataset. Here are our findings;

\begin{itemize}
    \item \textbf{Baby Dataset}: For this dataset we have determined that an optimal top K level is 35. In studies \cite{Zhang2021MiningLS}  we observed that this number was significantly lower. In our research by incorporating the circle loss function and the purifier layer we have enhanced the resilience of our model against data contamination. As a result we observed an increase, in the value of K.

    \item \textbf{Sports Dataset}: We noticed a scenario with this dataset. We found that the model benefits from having top K nearest neighbors. After analysis we have determined that the optimal level for top 30 is X.

    \item \textbf{Clothing Dataset}:  For this dataset, we have observed a rise in the number of top K closest neighbors that positively impact the models performance. Based on our analysis we have identified 30 as the level, for top K.
\end{itemize}

\begin{figure}[t]
    \centering
    --\caption{Top-K Sparsification: K Selection Effects}
    \includegraphics[width=1\linewidth]{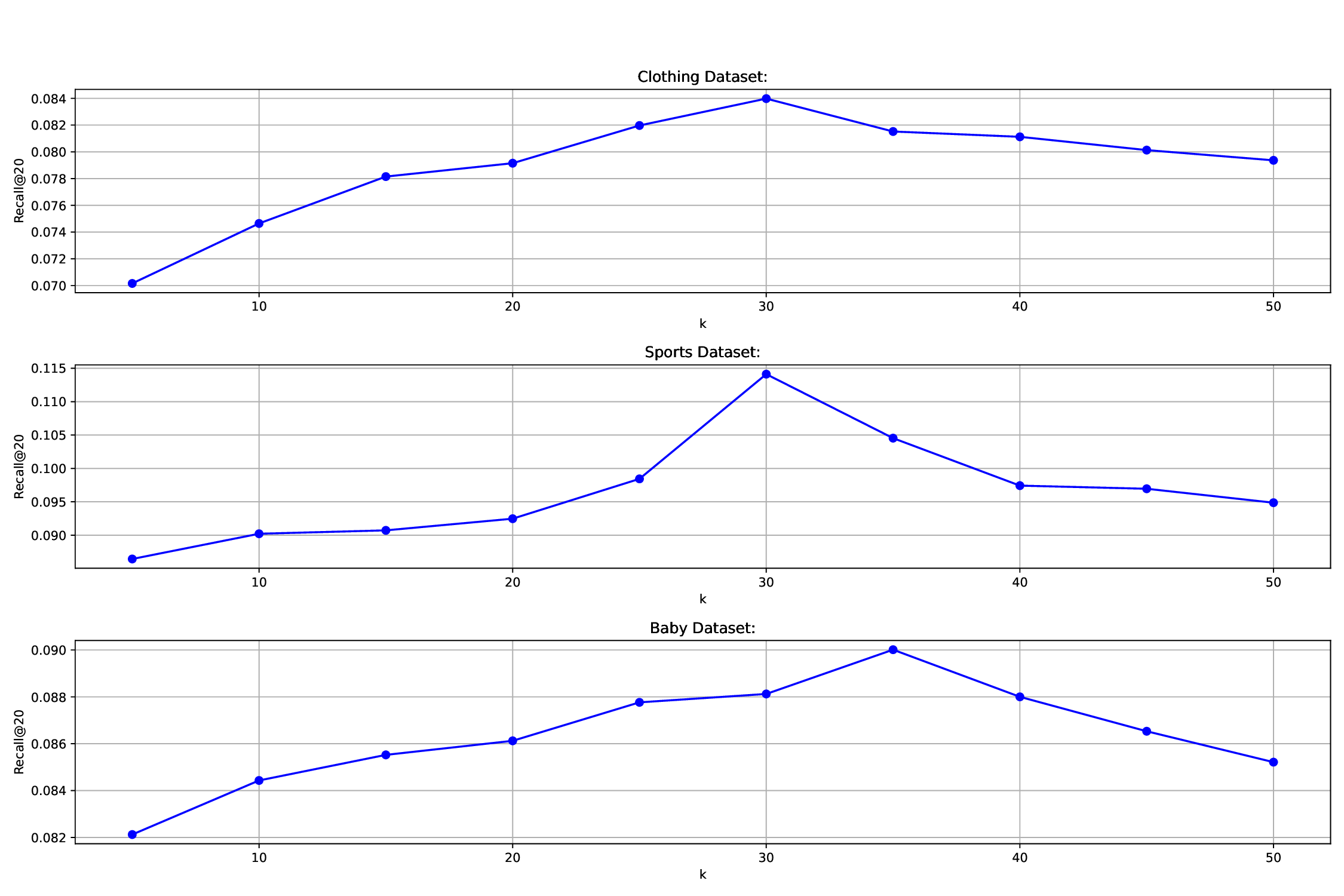}
\end{figure}

\subsection{Effects of Modalities}
This section of the document aims to explore how types of input information also known as modalities contribute to the performance of a system. It discusses Table 3 which presents a comparison of performance across modalities. The findings indicate that systems that incorporate modalities outperform those that rely on just one modality. This highlights the advantages of taking an approach when it comes to understanding items. The text emphasizes that textual information, which includes details like titles, categories and descriptions generally has a significant impact on performance compared to visual information. This is because textual data provides, in depth information while visual data tends to offer a broader representation of how items appear.

\begin{table}[ht]
\centering
\begin{tabular}{llcc}
\hline
\textbf{Dataset} & \textbf{Modality} & \textbf{Recall@20} & \textbf{NDCG@20} \\
\hline
Clothing & Visual & 0.0737 & 0.0348 \\
         & Textual & 0.0774 & 0.0361 \\
         & Both & \textbf{0.0840} & \textbf{0.0381} \\
\hline
Baby     & Visual & 0.0814 & 0.0345 \\
         & Textual & 0.0858 & 0.0380 \\
         & Both & \textbf{0.0900} & \textbf{0.0413} \\
\hline
Sports   & Visual & 0.0881 & 0.0374 \\
         & Textual & 0.0937 & 0.0391 \\
         & Both & \textbf{0.1036} & \textbf{0.0453} \\
\hline
\end{tabular}
\caption{The effects of modality on recommendation performance.}
\label{tab:modality_effects}
\end{table}

\subsection{Impact of Purifier, Circle Loss}
This text explores how different variations of models affect the efficiency of recommendation systems. It discusses important findings and gives a Table 4 as results;

Models that lack a modality purifier (referred to as w/o MP) don't perform well because directly integrating pre extracted features from different sources isn't effective for recommendation tasks. This is because there is a lot of noise in the data, which the modality purifier filters out keeping only the relevant preferences.

Signals that are not directly related to recommendations can have an impact on performance. It highlights the importance of leveraging relationships between items, which helps address the issue of sparse data.

Models that include a fused modalities perform better because they take into account variations in user preferences when purchasing different items. While these models are good, at utilizing user preferences they may overlook the significance of aspects or features of items resulting in an incomplete or biased representation and ultimately leading to suboptimal recommendation performance.

\begin{itemize}
    \item \textbf{w/o MP}: The modality purifier is not included in this approach. Instead the encoder directly receives the extracted features that are unique to each modality and processes the item, to item information.

    \item \textbf{w/o IIV}: The elimination of the item item information encoder streamlines the process by merging and propagating the features of each modality through a graph that represents user item interactions. This approach essentially simplifies the task to encoding collaboration signals.

    \item \textbf{w/o CIRCLE}: We remove circle loss.
\end{itemize}

\begin{table}[ht]
    \centering
    \begin{tabular}{llcc}
        \hline
        \textbf{Modules} & \textbf{Recall@20} \\
        \hline
        W/O MP & 0.0434 \\
        W/O IIV & 0.0841 \\
        W/O CIRCLE & 0.0937 \\
        Ours & 0.1036 \\
        \hline
    \end{tabular}
    \label{tab:effects_of_modules}
    \caption{Effects of Modules}
\end{table}

\subsection{Challenges and Limitations}

In this section we will discuss the challenges and limitations encountered in the study. We faced difficulties in pre processing and evaluating the data including a lack of open source datasets for benchmarking and the absence of an evaluation consortium. The topic of having a benchmark and evaluation consortium in recommendation systems is still evolving and remains controversial with no consensus reached at present.

The research heavily relies on circle loss, which serves a role. It helps manage noise within each modality while promoting information between embeddings in the fused modality by separating out noisy embeddings. By adjusting parameters specific to the dataset during training we can significantly enhance the models performance. However due to time constraints during model training and analysis conducting grid search, for hyperparameter tuning and circle loss parameters proved to be time consuming. Was left incomplete. Therefore we proceeded with the parameters chosen based on overall considerations.
\section{Conclusion and Future Works}

In this research we have introduced a method for recommending multimedia content focusing on a systematic process of filtering and improving various types of media. The approach starts with removing noise from forms of media then enhancing the refined features and behaviors separately using a specialized encoder that looks at item to item information. By combining a fusion technique with a self supervised auxiliary task we have gained a thorough understanding of user preferences paving the way for future improvements in recommendation systems.

The key aspect of our approach is its ability to extract and enrich the value of multimodal data providing a detailed view of user preferences and item attributes. By refining and purifying the features of each type of media before merging them we address the common problem of noise associated with different modalities ensuring that the combined features are relevant and informative. The use of an item to item encoder further enhances the systems capability to identify and utilize relationships, within the data improving the accuracy of recommendations.

Looking ahead our goal is to enhance our systems capacity to understand user preferences by incorporating domain expertise and utilizing language models.
This approach shows promise by tackling the core challenges of starting fresh in recommendation systems enhancing a more comprehensive understanding of user preferences and item characteristics within their context.

Our research has reaching implications providing a scalable and resilient framework for multimedia recommendation systems that can adjust and develop alongside the growing complexity and abundance of digital content. Not does our method set a new benchmark, for recommendation accuracy but it also paves the way for future studies to delve deeper into integrating multimodal data and sophisticated computational models propelling advancements in personalized recommendation systems.

In the future we aim to enhance the models ability to understand user preferences better by combining knowledge and utilizing advanced language models. This effort is focused on overcoming the obstacles in recommendation systems aiming for a deeper understanding of user preferences and item traits in their specific context. Exploring a range of approaches and delving into complex computational models represent key areas for upcoming research, which could potentially establish new standards, in personalized recommendation systems.

\bibliographystyle{plain}
\bibliography{main.bib}

\begin{thebibliography}{10}

\bibitem{Akiba2019OptunaAN}
Takuya Akiba, Shotaro Sano, Toshihiko Yanase, Takeru Ohta, and Masanori Koyama.
\newblock Optuna: A next-generation hyperparameter optimization framework.
\newblock {\em Proceedings of the 25th ACM SIGKDD International Conference on Knowledge Discovery \& Data Mining}, 2019.

\bibitem{Deshpande2004ItembasedTR}
Mukund Deshpande and George Karypis.
\newblock Item-based top-n recommendation algorithms.
\newblock volume~22, pages 143--177, 2004.

\bibitem{Feng2021EncoderFN}
Guang Feng, Zhiwei Hu, Lihe Zhang, and Huchuan Lu.
\newblock Encoder fusion network with co-attention embedding for referring image segmentation.
\newblock {\em 2021 IEEE/CVF Conference on Computer Vision and Pattern Recognition (CVPR)}, pages 15501--15510, 2021.

\bibitem{hamilton2017inductive}
William~L Hamilton, Rex Ying, and Jure Leskovec.
\newblock Inductive representation learning on large graphs.
\newblock {\em Advances in Neural Information Processing Systems (NeurIPS)}, pages 1024--1034, 2017.

\bibitem{He2015VBPRVB}
Ruining He and Julian McAuley.
\newblock Vbpr: Visual bayesian personalized ranking from implicit feedback.
\newblock In {\em AAAI Conference on Artificial Intelligence}, 2015.

\bibitem{he2020lightgcn}
Xiangnan He, Kuan Zhao, Xiang Xiao, Mengqing Xu, and Martin Ester.
\newblock Lightgcn: Simplifying and powering graph convolution network for recommendation.
\newblock {\em arXiv preprint arXiv:2002.02126}, 2020.

\bibitem{Hu2021MMGCNMF}
Jingwen Hu, Yuchen Liu, Jinming Zhao, and Qin Jin.
\newblock Mmgcn: Multimodal fusion via deep graph convolution network for emotion recognition in conversation.
\newblock {\em ArXiv}, abs/2107.06779, 2021.

\bibitem{hu2008collaborative}
Yifan Hu, Yehuda Koren, and Chris Volinsky.
\newblock Collaborative filtering for implicit feedback datasets.
\newblock In {\em 2008 Eighth IEEE International Conference on Data Mining}, pages 263--272. Ieee, 2008.

\bibitem{kipf2017semi}
Thomas~N Kipf and Max Welling.
\newblock Semi-supervised classification with graph convolutional networks.
\newblock In {\em International Conference on Learning Representations (ICLR)}, 2017.

\bibitem{koren2009}
Y.~Koren, R.~Bell, and C.~Volinsky.
\newblock Matrix factorization techniques for recommender systems.
\newblock {\em Computer}, 42(8), 2009.

\bibitem{lam2008addressing}
Xin Lam, Thanh Vu, Trung~D Le, and Anh~D Duong.
\newblock Addressing the cold-start problem in recommendation systems.
\newblock {\em PACIS 2008 Proceedings}, page 219, 2008.

\bibitem{linden2003amazon}
G.~Linden, B.~Smith, and J.~York.
\newblock Amazon. com recommendations: Item-to-item collaborative filtering.
\newblock {\em IEEE Internet computing}, 7(1):76--80, 2003.

\bibitem{Liu2023MultimodalRS}
Qidong Liu, Jiaxi Hu, Yutian Xiao, Jingtong Gao, and Xiang Zhao.
\newblock Multimodal recommender systems: A survey.
\newblock {\em ArXiv}, abs/2302.03883, 2023.

\bibitem{Loshchilov2017FixingWD}
Ilya Loshchilov and Frank Hutter.
\newblock Fixing weight decay regularization in adam.
\newblock {\em ArXiv}, abs/1711.05101, 2017.

\bibitem{malitesta2023formalizing}
Daniele Malitesta, Giandomenico Cornacchia, Claudio Pomo, Felice~Antonio Merra, Tommaso~Di Noia, and Eugenio~Di Sciascio.
\newblock Formalizing multimedia recommendation through multimodal deep learning, 2023.

\bibitem{amazon2014dataset}
Julian McAuley et~al.
\newblock Amazon product data, 2014.

\bibitem{McAuley2015ImageBasedRO}
Julian McAuley, Christopher Targett, Javen~Qinfeng Shi, and Anton van~den Hengel.
\newblock Image-based recommendations on styles and substitutes.
\newblock {\em Proceedings of the 38th International ACM SIGIR Conference on Research and Development in Information Retrieval}, 2015.

\bibitem{Reimers2019SentenceBERTSE}
Nils Reimers and Iryna Gurevych.
\newblock Sentence-bert: Sentence embeddings using siamese bert-networks.
\newblock In {\em Conference on Empirical Methods in Natural Language Processing}, 2019.

\bibitem{rendle2009}
S.~Rendle, C.~Freudenthaler, Z.~Gantner, and L.~Schmidt-Thieme.
\newblock Bpr: Bayesian personalized ranking from implicit feedback.
\newblock In {\em Proceedings of the Twenty-Fifth Conference on Uncertainty in Artificial Intelligence}, pages 452--461, 2009.

\bibitem{shi2015heterogeneous}
Chuan Shi, Zhiqiang Zhang, Ping Luo, Philip~S Yu, Yanchi Yue, and Bin Wu.
\newblock Heterogeneous information network embedding for recommendation.
\newblock {\em IEEE transactions on knowledge and data engineering}, 31(2):357--370, 2019.

\bibitem{Shi2019UnderstandingTS}
Shaohuai Shi, Xiaowen Chu, Ka~Chun Cheung, and S.~See.
\newblock Understanding top-k sparsification in distributed deep learning.
\newblock {\em ArXiv}, abs/1911.08772, 2019.

\bibitem{su2009survey}
Xiaoyuan Su and Taghi~M Khoshgoftaar.
\newblock A survey of collaborative filtering techniques.
\newblock {\em Advances in artificial intelligence}, 2009:4, 2009.

\bibitem{Sun2020CircleLA}
Yifan Sun, Changmao Cheng, Yuhan Zhang, Chi Zhang, Liang Zheng, Zhongdao Wang, and Yichen Wei.
\newblock Circle loss: A unified perspective of pair similarity optimization.
\newblock {\em 2020 IEEE/CVF Conference on Computer Vision and Pattern Recognition (CVPR)}, pages 6397--6406, 2020.

\bibitem{Wang2023DualGNNDG}
Qifan Wang, Yin wei Wei, Jianhua Yin, Jianlong Wu, Xuemeng Song, Liqiang Nie, and Min Zhang.
\newblock Dualgnn: Dual graph neural network for multimedia recommendation.
\newblock {\em IEEE Transactions on Multimedia}, 25:1074--1084, 2023.

\bibitem{wang2019neural}
Xiang Wang, Xiangnan He, Meng Cao, Meng Liu, and Tat-Seng Chua.
\newblock Neural graph collaborative filtering.
\newblock In {\em Proceedings of the 42nd International ACM SIGIR Conference on Research and Development in Information Retrieval}, pages 165--174, 2019.

\bibitem{Wei2023MultiModalSL}
Wei Wei, Chao Huang, Lianghao Xia, and Chuxu Zhang.
\newblock Multi-modal self-supervised learning for recommendation.
\newblock {\em Proceedings of the ACM Web Conference 2023}, 2023.

\bibitem{Wei2020GraphRefinedCN}
Yin wei Wei, Xiang Wang, Liqiang Nie, Xiangnan He, and Tat-Seng Chua.
\newblock Graph-refined convolutional network for multimedia recommendation with implicit feedback.
\newblock {\em Proceedings of the 28th ACM International Conference on Multimedia}, 2020.

\bibitem{Zhang2021MiningLS}
Jinghao Zhang, Yanqiao Zhu, Qiang Liu, Shu Wu, Shuhui Wang, and Liang Wang.
\newblock Mining latent structures for multimedia recommendation.
\newblock 2021.

\bibitem{zhou2018graph}
Jie Zhou, Ganqu Cui, Zhengyan Zhang, Cheng Yang, Zhiyuan Liu, Lifeng Wang, Changcheng Li, and Maosong Sun.
\newblock Graph neural networks: A review of methods and applications.
\newblock {\em AI Open}, 1:57--81, 2018.

\bibitem{Zhou2022BootstrapLR}
Xin Zhou, Hongyu Zhou, Yong Liu, Zhiwei Zeng, Chunyan Miao, P.~Wang, Yuan You, and Feijun Jiang.
\newblock Bootstrap latent representations for multi-modal recommendation.
\newblock {\em Proceedings of the ACM Web Conference 2023}, 2022.

\end{thebibliography}

\end{document}